# GoHammer Blockchain Performance Test Tool

**Melih Birim**
TUBU
İstanbul, Turkey
melih@tubu.io

**Hüseyin Emre Arı**
TUBU
İstanbul, Turkey
hemre@tubu.io

**Enis Karaarslan**
Department of Computer Enginnering
Muğla Sıtkı Koçman University
Muğla, Turkey
enis.karaarslan@mu.edu.tr

*Abstract*— Decentralized services are increasingly being developed and their Decentralized applications are increasingly developed but their performance metrics are not tested enough. The total number of transactions that can be supported by the blockchain network and the performance effects of selecting different consensus protocols, using different block intervals and block size should be tested. There are some blockchain performance tools but most are built for specific blockchain frameworks and require complex configuration. The GoHammer tool is developed to provide an easy to use, flexible test tool for the Ethereum/Quorum blockchain frameworks. Transaction per second (TPS) values and several performance metrics will be tested. This tool is also a part of the series of tools that can be integrated with Tubu-io. This tool will help in developing more efficient decentralized systems and will affect decreasing the costs of developing decentralized application projects.

## I. Introduction

Blockchain technology enables us to have trusted services without using an intermediary. Smart contracts that are introduced with Ethereum can be used to form autonomous Decentralized applications (DAPP) [1].

The availability and integrity of the blockchain systems increase with the number of nodes, but this also decreases the system performance [2]. However, performance tuning is possible by deploying different types of nodes and using different consensus protocols, using different block writing intervals or block sizes.

In the next section, performance evaluation in decentralized systems is explained. Related works are provided in the third section. The proposal of the system architecture and the implementation is explained in the fourth section. The results and conclusions are included in the final section.

## II. Performance Evaluation

The sample test scenario can be used for the performance evaluation is shown in Fig 1. It consists of the blockchain network which will be tested and the test node. The system needs at least one test device which will generate the load and then observe the results, which will be called the TestNode in the document. The main performance metrics are as follows:

- Latency:
  – Read Latency
  – Transaction Latency
- Throughput:
  – Read Throughput
  – Transaction Throughput
- Resource utilization (CPU, Memory, Network IO, . . .)
- Number of failed/delayed transactions due to timeouts [7]

## III. Related Works

There are several test tools for specific blockchain environments that are used to test decentralized applications (DAPP). Selected environments are compared in the Table 1. Currently, the most comprehensive performance benchmarking environment is the Hyperledger Caliper (https://hyperledger.github.io/caliper/). It supports many platforms other than Quorum. It can collect several performance metrics. Test transactions and scripts can be prepared for the use case scenarios. Sample implementations are given in [3-5, 7].

GoHammer tool is highly inspired by the Chainhammer (https://github.com/drandreaskrueger/chainhammer) and Consensys's Quorum Profiling [8]. We needed a test tool other than these two available tools as both are not easy to use and need complicated installation steps. Also Chainhammer's some required dependencies are outdated.

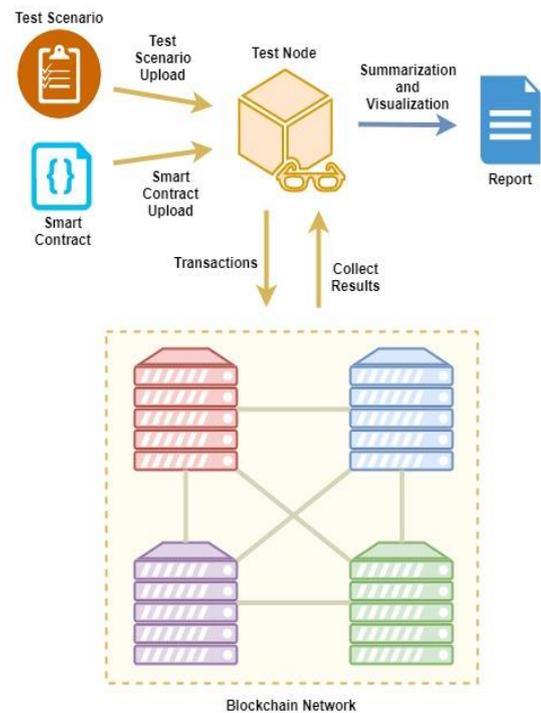

**Fig. 1:** Test Scenario

**Table 1.** Comparison table

| Product | *Caliper* | *Chainhammer* | *Quorum Profiling* | *GoHammer* |
|---|---|---|---|---|
| Platform | Hyperledger Besu, Ethereum, Hyperledger Fabric, FISCO BCOS | Quorum, Parity | Quorum | Quorum |
| Transaction/read throughput | Yes | Yes | Yes | Yes |
| Transaction/read latency | Yes | -[1] | No[1] | No |
| Resource consumption | Yes | No[1] | No | No |
| Consensus Protocol | -[1] | Raft, IBFT, Tobalaba | Raft, IBFT | Raft, IBFT[2] |
| Easy to Use | No | No | Partial | Yes |
| Minimal Installation [3] | No | No | Partial | Yes |
| Up to date | Yes | Partial | Partial | Yes |
| Programming Language | Javascript | Python | Golang | Golang |

[1] Not clearly specified, [2] In the next Release, [3] Easy to install and less number of dependencies

## IV. GoHammer Blockchain Performance Test System

GoHammer blockchain test tool generates transactions, runs test profiles and collects the TPS and resource consumption of the system. GoHammer is used with the TPS-Monitor components such as Grafana, InfluxDB and Telegraf. Telegraf is used to collect the data from the nodes. InfluxDB is used to store that data. Grafana is used to visualize the data. Docker is used for grafana and influxdb. The system architecture is shown in Fig 2.

GoHammer is developed using the golang programming language, especially for the Quorum blockchain framework. The test environment is formed of two main parts:

- Generating load on the system: Test profiles are managed and transactions are produced. The core GoHammer tool is used.
- Collecting and visualizing TPS and system metrics: TPS-Monitor (https://github.com/ConsenSys/quorumprofiling/tree/master/tps-monitor) that is taken from quorum-profiling toolset is used in this phase.

GoHammer parses the configuration file that contains information about the nodes involved and how the transactions will be produced. All transactions are produced on one node by default. If the scenario needs more than one Test Node, transactions will be produced with the round robin method. Consequent test profiles can be activated and each will have its own metrics and statistics. As an example; the first test scenario may have 1000 transactions and the second one 2000. The system will not generate any transaction for the given timeout value. Timeout values are set in human readable format like "1m", "30s", "10ms" etc. If you don't want any timeout you can set this value to "0s".

GoHammer has a compiled smart contract which can be used to produce transactions on the blockchain. The transactions can also be produced by calling this precompiled smart contract's "setItem" method by setting "callContractMethod" field to true in the test config file. Currently resource consumption is only collected when this tool is run on a blockchain node. Collecting for different blockchain nodes will be possible in future releases when an agent is used.

Gohammer summarizes the results after the test scenario is finished. The log file is automatically named with the timestamp of the test date and time (DD_MM_YY mm:hh.log). The test execution time, test section execution time and the number of transactions produced is included in the log. Visualization processes are implemented with Grafana and sample screenshots of the implementation are given in Fig 3 and Fig 4.

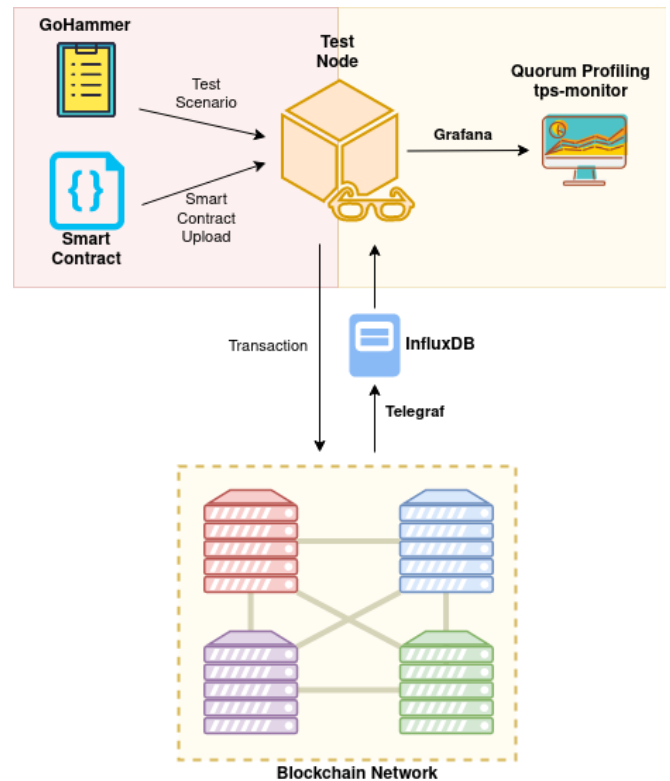

**Fig. 2:** Blockchain Test System Architecture

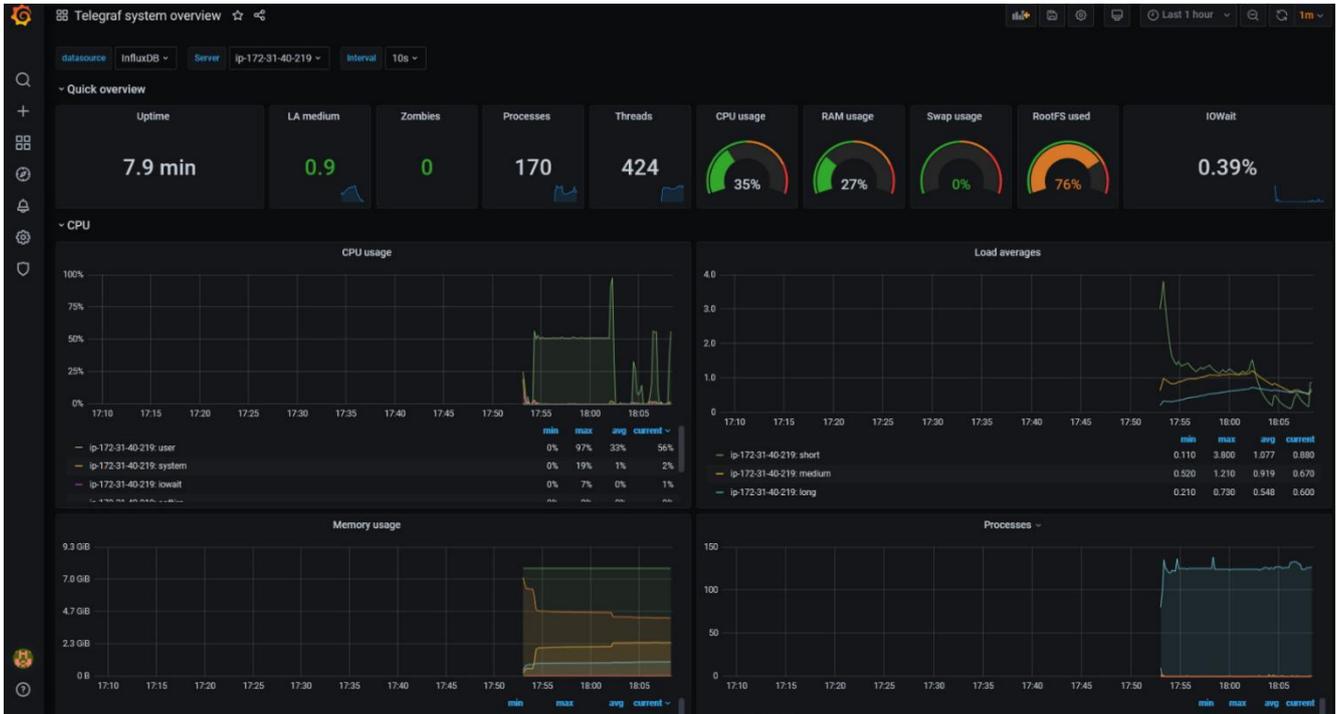

**Fig.3:** System Visualization Screen 1

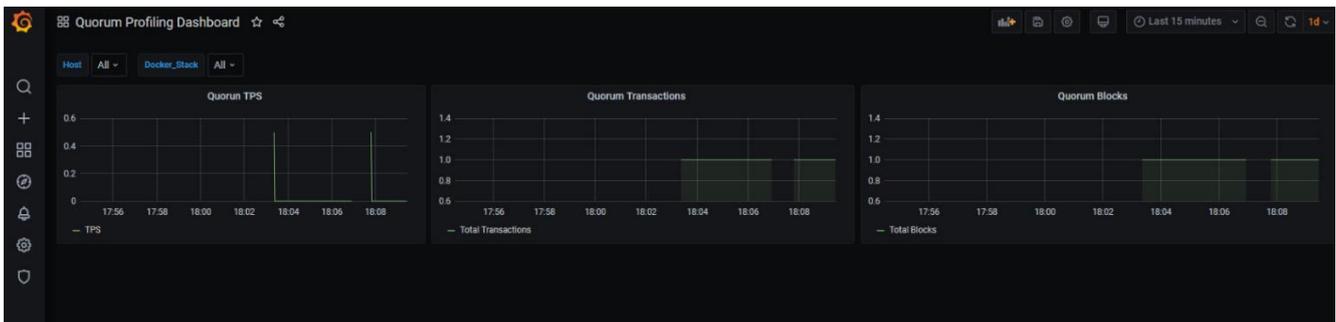

**Fig. 4:** System Visualization Screen 2

V. CONCLUSION

We use GoHammer not only for research and academic projects but also in our commercial projects to test our decentralized applications performance. We are still working to improve it by opening its source codes with GPL licence on GitHub (https://github.com/tubuarge/gohammer).

It is planned to integrate with Tubu-io [9]. It will help in developing more efficient decentralized systems and is supposed to affect decreasing the costs of developing decentralized projects.

ACKNOWLEDGMENT

We would also like to thank Cemal Dak for his contribution in the graphics.

REFERENCES

[1] Karaarslan, E. & Konacakli, E. (2020). Data Storage in the Decentralized World: Blockchain and Derivatives. In Gulsecen S., Sharma S., Akadal E.(Eds.), Who Runs The World: DATA (pp. 37-69). Istanbul, Istanbul University Press.

[2] Hyperledger Blockchain Performance Metrics (2018), Obtained from https://www.hyperledger.org/learn/publications/blockchainperformance-metrics

[3] Kuzlu, M., Pipattanasomporn, M., Gurses, L., Rahman, S. (2019, July). Performance analysis of a hyperledger fabric blockchain framework: throughput, latency and scalability. In 2019 IEEE international conference on blockchain (Blockchain) (pp. 536-540). IEEE.

[4] Pongnumkul, S., Siripanpornchana, C., Thajchayapong, S. (2017, July). Performance analysis of private blockchain platforms in varying workloads. In 2017 26th International Conference on Computer Communication and Networks (ICCCN) (pp. 1-6). IEEE.

[5] Nasir, Q., Qasse, I. A., Abu Talib, M., Nassif, A. B. (2018). Performance analysis of hyperledger fabric platforms. Security and Communication Networks, 2018.

[6] Hyperledger Getting Started, Obtained from https://hyperledger.github.io/caliper/v0.4.2/getting-started/

[7] Wickboldt, C. (2019). Benchmarking a blockchain-based certification storage system (No. 2019/5). Diskussionsbeiträge.

[8] Quorum Profiling (2021), Obtained from https://docs.goquorum.consensys.net/en/stable/Concepts/Profiling/

[9] Işık, E., Birim, M., Karaarslan, E. (2021). Tubu-io Decentralized Application Development Test Workbench. arXiv preprint arXiv:2103.11187.